\documentclass[12pt,a4paper,twoside]{article}
\usepackage[latin1]{inputenc}
\usepackage[english]{babel}
\usepackage[margin=3cm]{geometry}

\usepackage{hyperref}
\usepackage{graphicx}
\usepackage{times}
\usepackage{fancyhdr}
\usepackage{epsfig}
\usepackage{multirow}
\usepackage{amssymb,amsfonts,amsmath}

\newcommand{\ntag}[1]{{\sc #1}}

\usepackage[hang,small,bf]{caption}

\begin{document}

\renewcommand{\multirowsetup}{\centering}

\thispagestyle{empty} 

\begin{center}
 {\Large\bfseries\sffamily
 Analytical computation of the epidemic threshold on temporal networks\\
 }
 \vspace{1cm}\large
{Eugenio Valdano\textsuperscript{1,2},
Luca Ferreri\textsuperscript{3}
Chiara Poletto\textsuperscript{1,2},\\
and Vittoria Colizza\textsuperscript{1,2,4}}\\
\vspace{1.1cm}
{\footnotesize
\textsuperscript{a}INSERM, UMR-S 1136, Institut Pierre Louis d'Epid\'emiologie et de Sant\'e Publique, F-75013, Paris, France.\\
\textsuperscript{b}Sorbonne Universit\'es, UPMC Univ Paris 06, UMR-S 1136, Institut Pierre Louis d'Epid\'emiologie et de Sant\'e Publique, F-75013, Paris, France.\\
\textsuperscript{c}Dipartimento di Scienze Veterinarie, Universit\`a degli Studi di Torino, Torino, Italy.\\
\textsuperscript{d}ISI Foundation, Torino, Italy.\\
}
\end{center}

 \vspace{2.2cm}

{\sffamily\footnotesize
\centering{\bfseries Abstract}\\
The time variation of contacts in a networked system may fundamentally alter the properties of spreading processes and affect the condition for large-scale propagation, as encoded in the epidemic threshold. Despite the great interest in the problem for the physics, applied mathematics, computer science and epidemiology communities, a full theoretical understanding is still missing and currently limited to the cases where the time-scale separation holds between spreading and network dynamics or to specific temporal network models. We consider a Markov chain description of the Susceptible-Infectious-Susceptible process on an arbitrary temporal network. By adopting a multilayer perspective, we develop a general analytical derivation of the epidemic threshold in terms of the spectral radius of a matrix that encodes both network structure and disease dynamics. The accuracy of the approach is confirmed on a set of temporal models and empirical networks and against numerical results. In addition, we explore how the threshold changes when varying the overall time of observation of the temporal network, so as to provide insights on the optimal time window for data collection of empirical temporal networked systems. Our framework is both of fundamental and practical interest, as it offers novel understanding of the interplay between temporal networks and spreading dynamics.
}

\newpage


\section{Introduction}

A wide range of physical, social and biological phenomena can be expressed in terms of spreading processes on interconnected substrates. Notable examples include the spread of directly-transmitted infectious diseases through host-to-host contacts \cite{Keeling2008}, the spatial propagation of epidemics driven by the hosts' mobility network \cite{Keeling2008,Colizza2007,Tatem2010}, the spread of cyber viruses along computer connections~\cite{PastorSatorras2001}, or the diffusion of ideas mediated by social interactions \cite{Goffman1964,Daley1964}. These phenomena are the result of a complex interplay between the properties of the spreading dynamics and the network's structural and temporal features, hindering their full understanding.

A  fundamental issue characterizing spreading processes is the identification of the critical condition for the wide spreading regime, encoded in the epidemic threshold parameter.  This is of critical importance for epidemic containment \cite{Keeling2008}, as well as for control of the diffusion of information \cite{Bikhchandani1992} and cyber viruses~\cite{PastorSatorras2001}. Extensive studies have characterized this parameter in the timescale separation approximation, i.e. when the timescales of spreading process and network evolution  strongly differ. This includes the two limiting regimes, {\it quenched}  and {\it annealed}  \cite{PastorSatorras2001, Wang2003, Gomez2010, Cohen2000, Newman2002, Boguna2013, Castellano2010, Goltsev2012, Barrat2008}. In the first case, the network is regarded as static, as it evolves on much slower timescales than the ones characterizing the spreading process. The epidemic threshold in this case is computed from the adjacency matrix describing the network connectivity pattern \cite{Wang2003, Gomez2010}. In the second case, the underlying network evolves so rapidly with respect to the dynamical process that only its time-averaged properties are relevant to the spreading dynamics. Approaches like the heterogenous mean field \cite{PastorSatorras2001}, the generating function  \cite{Newman2002} and percolation theory  \cite{Cohen2000} provide, in this regime, estimates of the threshold in the infinite size limit. 

Recently, the extensive empirical characterization of social interactions at different scales and settings  \cite{Isella2011, Miritello2011, Rocha2010, Bajardi2011, Butts2009, Karsai2011, Holme2012} has shown that networks often display non-Poissonian and non-Markovian temporal evolutions unfolding at timescales similar to the ones of many spreading processes of interest, stressing the need for novel theoretical tools able to overcome current limitations. Much research has focused on spreading processes occurring on time-varying networks \cite{Isella2011,Miritello2011, Karsai2011, Rocha2013, Ferreri2014, Iribarren2009, Vazquez2007, Eames2004, Perra2012, Volz2009, Gross2006, Taylor2012}, modeled either as a discrete-time sequence of networks~\cite{Perra2012,Isella2011} or as continuous-time dynamics of links~\cite{Miritello2011, Vazquez2007}, however  only few studies have so far provided an analytical calculation of the epidemic threshold in specific cases~\cite{Ferreri2014,Eames2004, Perra2012, Liu2014, Volz2009, Gross2006, Taylor2012}. These are all based on models for time-varying networks integrating the microscopical laws governing the network evolution, under context-specific assumptions. An analytical framework for the computation of the epidemic threshold for an arbitrary  time-varying network is still missing. To fill such gap, we present here a novel approach that, by reinterpreting the tensor formalism of multi-layer networks \cite{DeDomenico2013,Kivela2014}, extends the Markov chain approach adopted for static networks \cite{Wang2003, Gomez2010} to their temporal counterpart. The approach is applied to discrete time-varying network models and empirical networks to highlight the role of different dynamical features on the spreading potential. The role of the observation time window is then analyzed in depth in order to provide indications on how this factor alters the estimated epidemic threshold.

\section{Derivation of the epidemic threshold}

We consider the Susceptible-Infected-Susceptible (SIS) model~\cite{Keeling2008} in discrete time, where individuals (i.e. the nodes of the network) can be in one of two mutually exclusive states~--~susceptible or infectious. At each time step, infectious individuals may transmit the infection to susceptible neighbors with probability $\lambda$ along each contact, and they recover spontaneously with probability $\mu$ becoming  susceptible once again. We consider the temporal network forming the substrate of the spreading process to be a sequence of undirected and unweighted static networks. The generalization of the following treatment to the directed and weighted case is outlined in the Supplementary Information.

In order to describe the spreading dynamics on such a substrate we extend  the Markov chain approach for static networks~\cite{Wang2003,Gomez2010} to the case of temporal networks. The SIS propagation on a generic network with $N$ nodes and adjacency matrix $A$ is given by:
\begin{equation}
  p_i^{(t)} = 1-\left[ 1- \left(1-\mu\right)p_i^{(t-1)} \right]  \prod_j \left[ 1-\lambda A_{ji} p_j^{(t-1)} \right],
 \label{eq:basic1Gen}
\end{equation}
where $p_i^{(t)}$ is the probability for the node $i$ to be in the infectious state at time $t$. The Markovian model of Eq.~(\ref{eq:basic1Gen}), widely adopted in different fields~\cite{Boguna2013,VanMieghem2011},
is based on the  mean-field assumption of the absence of dynamical correlations among the states of neighboring nodes \cite{Gleeson2012}. For both directed and undirected networks \cite{VanMieghem2013, Li2013} the study of the asymptotic state yields the derivation of the epidemic threshold $\left(\lambda/\mu \right) = 1/\rho(A^\dagger)$, where $\rho (A^\dagger)$  is the spectral radius of the transposed adjacency matrix $A^\dagger$ \cite{Wang2003, Gomez2010}. This is known to be a lower bound estimate of the real epidemic threshold, approaching the real value with surprising high accuracy given the simplicity of the expression and its derivation \cite{Gleeson2012, Cator2012}. 

We extend this paradigm to a temporal network by letting the adjacency matrix in Eq.~(\ref{eq:basic1Gen}) depend on time:
\begin{equation}
 p_i^{(t)} = 1-\left[ 1- \left(1-\mu\right)p_i^{(t-1)} \right]  \prod_j \left[ 1-\lambda A_{ji}^{(t-1)} p_j^{(t-1)} \right].
 \label{eq:temporal}
\end{equation}
Here $A^{(t)}$ is the adjacency matrix associated to the $t$-th snapshot of the evolving network. 
In order to ensure the asymptotic solution of the SIS process in a generic temporal network we assume periodic boundary conditions for the network dynamics.  Being $T$ the total number of network time snapshots, we impose $A^{(T+1)}\equiv A^{(1)}$. This does not imply any loss of generality given that $T$ may be completely arbitrary. We notice that, as a consequence of the assumed periodic temporal dynamics of $A^{(t)}$, the asymptotic solution of Eq.~(\ref{eq:temporal}) is in principle periodic, with period $T$.

We now define  a more convenient representation of the coupled dynamics adopting the multi-layer approach introduced in \cite{DeDomenico2013}. We map the temporal network to the tensor space $\mathbb{R}^{N}\otimes \mathbb{R}^T$, where each node is identified by the pair of indices $(i,t)$, corresponding  to the node label, $i$, and the time frame, $t$, respectively. A multi-layer representation of the temporal network can be introduced through the following rules:
\begin{itemize}
 \item each node, at time $t$, is connected to its future self at $t+1$;
 \item if $i$ is connected to $j$ at time $t$, then we connect $i$ at time $t$ to $j$ at time $t+1$, and, $j$ at time $t$ to $i$ at time $t+1$.
\end{itemize}
The second rule is termed non-diagonal coupling in the multilayer-network framework~\cite{Kivela2014}.
The first rule is consistent with the ordinal coupling in such framework~\cite{Kivela2014,Mucha2010}, but unlike in that representation, no links are found connecting nodes on the same layer, since layers cease to correspond to the adjacency matrices of the temporal snapshots. The so-defined network is therefore multipartite, since only pairs of nodes belonging to different layers are linked together (see Fig.~\ref{fig:schema} for a schematic illustration of this transformation). While formally falling into a specific sub-case of the classification introduced in~\cite{Wehmuth2014}, the proposed mapping from the network temporal sequence to a multilayer object provides a novel representation of the temporal network that preserves the information relevant for the spreading process.
The tensor representation of the obtained multilayer network is the following:
\begin{equation}
\mathbf{A}_{ij}^{tt'} = \delta^{t,t'+1}\left[  \delta_{ij} + A^{(t)}_{ij}  \right].
 \label{eq:matr_multipartite-netonly}
\end{equation}
Analogously to the definition of $\mathbf{A}$, we can also write in this representation the tensor associated to the SIS dynamics of Eq.~(\ref{eq:temporal}),  coupling together contagion and network dynamics:
\begin{equation}
 \mathbf{M}_{ij}^{tt'} = \delta^{t,t'+1}\left[  \left(1-\mu\right)\delta_{ij} + \lambda A^{(t)}_{ij}  \right].
 \label{eq:matr_multipartite}
\end{equation}
The multi-layer representation and the definition of the tensor $\mathbf{M}$ introduces a simplified expression for Eq.~(\ref{eq:temporal}). The tensor space can be represented in single index notation through the isomorphism $\mathbb{R}^{N}\otimes \mathbb{R}^T\simeq \mathbb{R}^{NT}$. In other words, similarly to the definition of the supra-adjacency matrix in \cite{DeDomenico2013, Cozzo2013b,Wang2013}, we can mask the tensorial origin of the space through the map $(i,t) \to \alpha=  Nt+i$, with $\alpha$ running in $\left\{ 1, ... , NT \right\}$, allowing us to write the network tensor $\mathbf M$ in matrix form 
\[
M = 
\left(
\begin{smallmatrix} 
0 & 1-\mu+\lambda A^{(1)} & 0 & \cdots & 0 \\
0 & 0 & 1-\mu+\lambda A^{(2)} & \cdots & 0 \\
\vdots & \vdots & \vdots & \vdots & \vdots \\
0 & 0 & 0 & \cdots & 1-\mu+\lambda A^{(T-1)} \\
1-\mu+\lambda A^{(T)} & 0 & 0 & \cdots & 0 \\
\end{smallmatrix} 
\right).
\]
$M$ provides a network representation of the topological and temporal dimensions underlying the dynamics of Eq.~(\ref{eq:temporal}), that here are interrelated and flattened. Its directed nature preserves the causality of the process, while its weights account for the SIS transition probabilities. The Markov process is now described by a trajectory in $\mathbb{R}^{NT}$ where the state vector $\hat{p}_\alpha(\tau)$ represents the probability of each node to be infected at each time step $t$ included in the interval $\left[\tau T, (\tau +1) T \right]$. Consistently, Eq.~(\ref{eq:temporal}) becomes 
\begin{equation}
 \hat{p}_\alpha (\tau) = 1-  \prod_\beta \left[ 1-  M_{\beta \alpha} \hat{p}_\beta \right (\tau-1) ].
 \label{eq:temporal2}
\end{equation}
Given that vector $\hat{p}$ encodes a $1$-period configuration, the $T$-periodic asymptotic state of the SIS process is now mapped into the steady state $\hat{p}_\alpha(\tau)=\hat{p}_\alpha(\tau-1)$. The latter can be recovered as solution of the equilibrium equation:
\begin{equation}
 \hat{p}_\alpha = 1-  \prod_\beta \left( 1-  M_{\beta\alpha} \hat{p}_\beta \right),
 \label{eq:temporal3}
\end{equation}
that is formally the same as the stationary condition imposed on the Eq.~(\ref{eq:basic1Gen}) for the static network case, and is similar to the Markov chain approaches used to solve contagion processes in multiplex and interconnected networks~\cite{Cozzo2013b,Wang2013,Granell2013}. We can then follow~\cite{Wang2003,Gomez2010} and linearize Eq.~(\ref{eq:temporal3}) recovering the necessary and sufficient condition for the asymptotically stable zero solution, $\rho\left(M^\dagger\right)<1$~\cite{Elaydi}. Considering that the uniform zero solution in the $\mathbb{R}^{NT}$ representation is mapped to a uniform zero solution in  the original $\mathbb{R}^N$ representation, this yields the threshold condition 
\begin{equation}
\rho\left( M^\dagger \right) = 1
 \label{eq:threshold}
\end{equation}
for the critical values of $\lambda$ and $\mu$ above which the the transmission becomes epidemic~\cite{Wang2003, Gomez2010,Cozzo2013b,Wang2013,Granell2013}. 

The spectral radius of $M$ can be simplified with the following relation (see Methods section):
\begin{equation}
 \rho\left( M \right) =  \rho\left( P \right)^{1/T}
 \label{eq:detM}
\end{equation}
where $P = \prod_{t=1}^T \left( 1-\mu+\lambda A^{(T-t)} \right)$ represents a weighted version of the accessibility matrix~\cite{Lentz2013}, having connectivity weighted by $\lambda$ and waiting time weighted by $1-\mu$. This last passage ensures a simplification of the numerical computation of the epidemic threshold, allowing an execution time scaling as $\sim TN^{5/2}$ (see the SI for an analysis of the numerical performance of our approach).

The quenched and annealed regimes can be recovered within this  general framework as particular limiting solutions. In the first case, 
it is to be noted that the sequence of temporal snapshots naturally defines the minimum timescale of the process. In order to consider contagion dynamics that are much faster than the time-varying process of the network, 
we thus rely on the commonly adopted assumption regarding the temporal network as static, so that $A^{(t)}\equiv A$.
In this particular case, $P=\left( 1-\mu + \lambda A \right)^T $. Therefore $\rho (M)=\rho(1-\mu + \lambda A ) = 1-\mu+\lambda\rho(A) $. The requirement $\rho(M^\dagger)=1$ thus recovers the expression known for the quenched case.

The study of the annealed regime is less trivial. In the assumption that $\lambda$ and $\mu$ are very small, corresponding to a very slow disease dynamics with respect to the timescale of the network evolution, it is possible to replace $P$ with its linear expansion in $\lambda/(1-\mu)$, yielding 
\[
 P_{\text{slow}} = \left(1-\mu \right)^T \left[   1+ \frac{\lambda}{1-\mu} {\mathcal A } \right],
\]
where
$
 {\mathcal A } = \sum_t A^{(t)}
$ is a weighted static representation of the network, formed by the sum of all the snapshots. Temporal correlations are lost, and edges count  for the number of times they are active during the whole period $T$. Eq.~(\ref{eq:threshold}) for the epidemic threshold thus simplifies to $\left( \lambda/\mu \right)_{\text{c,slow}} = T/\rho(\mathcal A)$ \cite{Masuda2013}, and the aggregated matrix contains all the relevant information for spreading dynamics.

\section{Validation and comparison with stochastic simulations}

In the following we validate the analytical method and compare its predictions with the behavior of a simulated SIS spreading process. For this purpose we consider six networks, three of them  built from  models for time-varying networks and the others obtained from empirical measures. The first network, \ntag{er}, is formed by a sequence of random Erd\H{o}s-R\'{e}nyi graphs \cite{Erdos1960} with a given number of nodes and edges. It represents a simple and completely uncorrelated example of temporal network. The second network, \ntag{activity},  is a realization of the activity driven model \cite{Perra2012} where each node is assigned an activity potential, representing the probability of being active in a certain snapshot. Once activated, the node establishes a fixed number of connections that are renewed at each snapshot. We consider a heterogeneous activity distribution, so that the obtained networks are characterized by a temporally uncorrelated sequence of snapshots with a heterogeneous aggregated degree distribution.  \ntag{bursty} is built from the model introduced in \cite{Rocha2013} and accounts for a heterogenous activation pattern describing a sequence of homogeneous networks where the inter-contact time is power-law distributed. Size and period are chosen arbitrarily for all these networks, since the choice of these parameters does not impact the method validation, as discussed more in detail in the following section. As examples of real time-varying networks, we choose datasets describing human contacts  of different kind: \ntag{ht09} is the network of face-to-face proximity during a 2.5 days scientific conference \cite{Isella2011};  \ntag{sex} is a 1-year network of sexual contacts between prostitutes and their clients \cite{Rocha2010}; \ntag{school} is a contact network describing one day in a high-school \cite{Salathe2010}. Size, period and topological properties are constrained by the measurements and are very diverse. Further information about the six networks can be found in the Appendix~\ref{ap:networks} and in Table~\ref{tab:networks}.

To verify the validity of the proposed analytical expression, we numerically solve  the Markov Eq.~(\ref{eq:temporal}). For given $\lambda$ and $\mu$, we iterate the equation until the periodic state is reached and  compute the average prevalence over a period $\left \langle i_{\text{MC}} \right\rangle = \sum_{i,t} p^{(t)}_i/(TN)$. Predictions are then also compared with the threshold behavior obtained from numerical simulations of the stochastic and microscopic SIS dynamics on the evolving networks. We use the quasi-stationary state method \cite{Ferreira2012} (see Methods section) to measure  the average prevalence $\langle i_{\text{sim}} \rangle$ over the time-series for different values of $\lambda$, after an initial transient time is discarded.

Fig.~\ref{fig:validation} shows  $\langle i_{\text{MC}} \rangle$ and $\langle i_{\text{sim}} \rangle$ as functions of $\lambda$ for two different values of $\mu$ ($\mu=0.2$ and $\mu=0.5$) for all networks under study. The average prevalence displays the expected transition behavior. The solution of the Markov chain equation $\left\langle i_{\text{MC}} \right\rangle$ is equal to zero for small values of $\lambda$ until the critical value of $\lambda$ is reached, after which a rapid growth is observed signaling an epidemic affecting a finite fraction of the network. The transition is well predicted by the analytical expression of Eq.~(\ref{eq:threshold}). The threshold behavior obtained from numerical simulations is also very similar to the mean-field prediction. The two curves of $\langle i_{\text{MC}} \rangle$ and $\langle i_{\text{sim}} \rangle$ are nearly superimposed, showing that the mean-field approximation in Eq.~(\ref{eq:temporal}) is valid in all conditions of network size and average connectivity here considered. The presence of correlations shows its effects in proximity of the transition that is smoother for  $\langle i_{\text{sim}} \rangle$  with respect to $\langle i_{\text{MC}} \rangle$. This is particularly evident for the network \ntag{ht09} and  is a consequence of its small size ($N=113$). In Supplementary Information we report the analysis of the dynamical correlations. 

The good agreement between the computed epidemic threshold, the solution of the Markov chain equation and the numerical simulation results is thus maintained under a range of different temporal network properties (presence or absence of temporal correlations, heterogeneous vs. homogeneous distributions characterizing temporal and structural observables and possible presence of community structure as in the case of the school) and sizes (from approximately $10^2$ nodes to $10^4$). It is important to mention that periodic boundary conditions in the case of real networks may in principle induce non-existing phenomena (such as, for example, temporal paths~\cite{RajKumar2011}) that could alter the threshold estimation by influencing the spreading process. We analyze the effect of our technical assumption of adopting periodic constraints  in the following section, also in relation to data availability and collection.   

\section{Optimal data collection time}

Available datasets characterizing empirical networks only account for a portion of the real contact process, and the extent of the recording time window may affect the prediction of the epidemic threshold. One may expect that, when the data-collection period is long enough, the data would represent an approximately complete reconstruction of the temporal network properties, thus leading to an accurate estimate of the epidemic threshold. Given the resources needed for the setup of data collection deployments, we explore here the role of the period $T$ aimed at identifying a minimum length of observation of the contact process that is optimal in providing a reliable characterization of the spreading potential.

We thus compute the epidemic threshold from Eq.~(\ref{eq:detM}) for increasingly larger values of the period $T$ up to the entire data-collection time window, for the three empirical networks under study. Fig.~\ref{fig:betac_T} shows a saturation behavior for $\lambda_c$, indicating that the data-collection period is long enough to characterise the epidemic dynamics.
Such behavior and its associated relaxation time strongly depend on the network typical timescale and on the temporal variability of its structure. More in detail, a simple structural measure~--~the variation of the average degree along the period~--~is shown to strongly impact the predicted $\lambda_c$ (Fig.~\ref{fig:betac_T} and SI).
This is particularly evident in the \ntag{school} network, where the daily activity of students determines considerable variations in the average degree and induces marked oscillations in the resulting threshold. Next to the empirical networks, we also consider the \ntag{bursty} network model as it includes non-trivial temporal correlations. In this case, the critical transmission probability $\lambda_c$ rapidly saturates to a constant value (Fig.~\ref{fig:betac_T}), and an even more rapid saturation is observed for the other two network models (see SI). The average degree is indeed relatively stable, so that small temporal windows are enough to fully characterize $\lambda_c$. 

Different values of the recovery probability $\mu$ lead in general to similar behaviors of $\lambda_c$ towards saturation, differing essentially by a scaling factor. The effect of $\mu$ on saturation time is instead visible for the \ntag{bursty} network. In this case, when the period length is smaller than the average duration of the infection, the truncation in the inter-contact time distribution clearly alters the estimation of the threshold.

These results indicate that it is possible to identify a minimum length of the observation window of a real system for contact data collection, highlighting the presence of well-defined properties and patterns characterizing the system that can be captured in a finite time.

\section{Conclusion}
Being able to provide a reliable and accurate estimation of the epidemic threshold for a spreading process taking place on a given networked system is of the utmost importance, as it allows predictions of the likelihood of a wide-spreading event and identification of containment measures (crucial for infectious disease epidemics) or strategies for enhancing the propagation (desired in the case of information diffusion). While analytical approaches have so far targeted only specific contexts, our framework allows the analytical computation of the epidemic threshold on an arbitrary  temporal network, requiring no assumption about the network topology or time variation. The proposed approach is based on the spectral decomposition of the flattened matrix representation of the topological and temporal structure of the network, extending the Markov chain model introduced for the static network case to its temporal counterpart. The predicted epidemic threshold, validated against the numerical solution of the model also reproduces the behavior observed in stochastic microscopic numerical simulations of the spreading process with high accuracy.

The technical requirement of periodic conditions does not limit the general applicability of our approach, as the method is valid for an arbitrary period length. Moreover, this feature allows us to inform data collection endeavors on the time period of observation of the system required to fully characterize its spreading properties. Our focus on the discrete time formulation of the process is prompted by the study of several empirical networks for which time is naturally discrete, the time step being dictated by the resolution of the data collection procedure. Extensions to the continuous-time case would be needed when the continuous time description is more appropriate, as for example with some modeling approaches.

Our framework thus introduces a multilayer formulation of spreading phenomena on time-varying networks that opens the path to new theoretical understandings of the complex interplay between the two temporal processes, disentangling the role of the network dynamical features, such as activation rate, temporal correlations and temporal resolution. 

Python code for computing the epidemic threshold can be found on\\ \texttt{https://github.com/eugenio-valdano/threshold}.

\appendix

\section{Proof of Eq.~(\ref{eq:detM})}

Computing the eigenvalues of $M^\dagger$ means solving the equation $\det \left(x-M^\dagger\right)=0$, where the determinant is computed on the $\mathbb{R}^{NT}$ space ($\det_{NT}$). Given that $x-M^\dagger$ is composed of $T^2$ blocks of size $N\times N$, we can use the findings in \cite{powell2011} to reduce the dimensionality of the problem, i.e. $\det_{NT}\rightarrow\det_{N}$. Moreover, given that several blocks of $x-M^\dagger$ are zero, the general result in \cite{powell2011} simplifies to
$ \det_{NT}\left( x- M^\dagger \right) = \left(-1\right)^{NT} \det_{N}\left( x^T- P \right) $. Eq.~(\ref{eq:detM}) immediately follows.

\section{Networks considered for validation} \label{ap:networks}

In this section we provide the details of the six networks used for the validation of the threshold expression.

\noindent
\ntag{er}: The network is formed by a sequence of random Erd\H{o}s-R\'{e}nyi graphs \cite{Erdos1960} with 500 nodes and 750 edges, so that $\left\langle k \right\rangle = 3$. 

\noindent
\ntag{activity}: In the activity driven model by Perra et al.~\cite{Perra2012}, nodes that are active in a certain snapshot establish a fixed number of connections (in our case $2$) with other nodes picked at random (both active and inactive). All links are renewed after every snapshot. The activity potential is assigned by sampling numbers $x\in[\epsilon,1]$ from a power-law distribution (in our case with exponent $2.8$ and $\epsilon=3\cdot 10^{-2}$), and then converting them to activity potentials $a=1-e^{-\eta x}$. $\eta$ is a free parameter used to tune the average degree,  here $\eta=10$.

\noindent
\ntag{bursty}: The network is built from the model introduced by Rocha et al.~\cite{Rocha2013} where a node becomes active at time $t$ with a probability that depends on the time it was last active, $t'$. If active, it then forms a link with another active node. All links are removed before proceeding to the following snapshots. To enforce a bursty inter-event time distribution, the probability of becoming active is sampled with the distribution $(t-t')^{-\alpha_1} e^{-\alpha_2 (t-t')}$, with $\alpha_1=2$ and $\alpha_2=5\cdot 10^{-4}$.

\noindent
\ntag{ht09}: The dataset was collected by the Sociopatterns group \cite{Isella2011}, and records the interactions among participants at a scientific conference. Links represent face-to-face proximity recorded by wearable RFID (Radio Frequency IDentification) tags. Time resolution of the signal is 20 sec. Each tag emits wave packets that may be recorded by other tags, thus signaling proximity. Tags were embedded in conference badges, their signal intensity was set to be detectable at a maximum distance of $1m$ and completely shielded by the human body. With these settings in effect, only close proximity in a face-to-face position resulted in interaction. Datasets and additional information are available on \texttt{www.sociopatterns.org}.

\noindent
\ntag{sex}: It is a network of sexual contacts between female prostitutes and their male clients as inferred from posts on a Brazilian online escort forum where customers could rate their experience with a certain sex worker. The date of the post was taken as a proxy for the time of the sexual intercourse. The (anonymized) dataset can be found in \cite{Rocha2010}.

\noindent
\ntag{school}: This network represents the face-to-face proximity interactions during a day in a high school \cite{Salathe2010}. Students and staff were given wearable RFID sensors, and proximity was recorded in a similar fashion as for \ntag{ht09}. Data can be found on \texttt{www.salathegroup.com}.

\section{Estimation of the epidemic threshold from numerical simulations}

The computation of  $\langle i_{\text{sim}} \rangle$ in proximity of the transition is made difficult because surviving configurations are rare and a very large number of realizations of the process is needed to collect substantial statistics. We use the quasistationary state (QS) method \cite{Ferreira2012, Binder2010}  to overcome this difficulty and increase our computational efficiency. The QS method is based on the idea of constraining the system in an active state. Every time the absorbing state of zero infected is reached by the system, it is substituted with an active configuration that is randomly taken from the history of the simulation. In particular, 50 active configurations for each network snapshot are kept in memory. Whenever an active configuration is reached, it replaces one of the 50 with probability 0.2. When the absorbing state is reached, an active configuration is chosen among these 50 of that particular snapshot. For each simulation, after a relaxation time of $3\cdot 10^3$ time steps, statistics are collected during $10^5$ time steps.
The method produces a time series that is long enough to accurately compute the observables $\langle i_{\text{sim}} \rangle$. 

\vspace{2cm}
\noindent
{\bf Acknowledgments}
\noindent
We thank Ciro Cattuto, Mario Giacobini, Sandro Meloni, Yamir Moreno, Nicola Perra, Mason Porter and Alessandro Vespignani for fruitful discussions. This work has been partially funded by the EC-Health contract no. 278433 (PREDEMICS) to VC and CP; the ANR contract no. ANR- 12-MONU-0018 (HARMSFLU) to V.C.; the PHC Programme Galilee contract no. 28144NH, the Italian Ministry of Health contract no. IZS AM 04/11 RC, the EC-ANIHWA contract no. ANR-13-ANWA-0007-03 (LIVEepi) to  E.V., C.P., V.C.; the ``Pierre Louis'' School of Public Health of UPMC, Paris, France to E.V.; the \href{http://www.progettolagrange.it/en}{Lagrange Project on Complex Systems} -- CRT and ISI Foundation to L.F..

\newpage


\bibliography{valdano_arXiv_definitive}{}

\begin{thebibliography}{10}

\bibitem{Keeling2008}
Matthew~James Keeling and Pejman Rohani.
\newblock {\em {Modeling infectious diseases in humans and animals}}.
\newblock Princeton Univ Pr, 2008.

\bibitem{Colizza2007}
Vittoria Colizza, Alain Barrat, Marc Barthelemy, Alain-Jacques Valleron, and
  Alessandro Vespignani.
\newblock Modeling the worldwide spread of pandemic influenza: Baseline case
  and containment interventions.
\newblock {\em PLoS Med}, 4(1):e13, 2007.

\bibitem{Tatem2010}
Andrew~J. Tatem and David~L. Smith.
\newblock International population movements and regional plasmodium falciparum
  malaria elimination strategies.
\newblock {\em Proceedings of the National Academy of Sciences},
  107(27):12222--12227, 2010.

\bibitem{PastorSatorras2001}
Romualdo Pastor-Satorras and Alessandro Vespignani.
\newblock {Epidemic spreading in scale-free networks.}
\newblock {\em Phys. Rev. Lett.}, 86(14):3200--3203, 2001.

\bibitem{Goffman1964}
W~Goffman and VA~Newill.
\newblock {Generalization of Epidemic Theory: An Application to the
  Transmission of Ideas}.
\newblock {\em Nature}, 204(4955):225--228, 1964.

\bibitem{Daley1964}
D~J Daley and D~G Kendall.
\newblock {Epidemics and Rumours}.
\newblock {\em Nature}, 204(4963):1118, 1964.

\bibitem{Bikhchandani1992}
Sushil Bikhchandani, David Hirshleifer, and Ivo Welch.
\newblock {A Theory of Fads, Fashion, Custom, and Cultural Change in
  Informational Cascades}.
\newblock {\em Journal of Political Economy}, 100(5):992--1026, 1992.

\bibitem{Wang2003}
Yang Wang, D.~Chakrabarti, Chenxi Wang, and C.~Faloutsos.
\newblock Epidemic spreading in real networks: an eigenvalue viewpoint.
\newblock In {\em Reliable Distributed Systems, 2003. Proceedings. 22nd
  International Symposium on}, pages 25--34, 2003.

\bibitem{Gomez2010}
Sergio G{\'o}mez, Alexandre Arenas, J~Borge-Holthoefer, Sandro Meloni, and
  Yamir Moreno.
\newblock {Discrete time Markov chain approach to contact-based disease
  spreading in complex networks}.
\newblock {\em EPL (Europhysics Letters)}, 89(3):38009, 2010.

\bibitem{Cohen2000}
Reuven Cohen, Keren Erez, Daniel ben Avraham, and Shlomo Havlin.
\newblock Resilience of the internet to random breakdowns.
\newblock {\em Phys. Rev. Lett.}, 85:4626--4628, 2000.

\bibitem{Newman2002}
M.~E.~J. Newman.
\newblock Spread of epidemic disease on networks.
\newblock {\em Phys. Rev. E}, 66:016128, 2002.

\bibitem{Boguna2013}
Marian Bogu\~n\'a, Claudio Castellano, and Romualdo Pastor-Satorras.
\newblock Nature of the epidemic threshold for the
  susceptible-infected-susceptible dynamics in networks.
\newblock {\em Phys. Rev. Lett.}, 111:068701, 2013.

\bibitem{Castellano2010}
Claudio Castellano and Romualdo Pastor-Satorras.
\newblock {Thresholds for Epidemic Spreading in Networks}.
\newblock {\em Phys. Rev. Lett.}, 105:218701, 2010.

\bibitem{Goltsev2012}
A.~V. Goltsev, S.~N. Dorogovtsev, J.~G. Oliveira, and J.~F.~F. Mendes.
\newblock Localization and spreading of diseases in complex networks.
\newblock {\em Phys. Rev. Lett.}, 109:128702, 2012.

\bibitem{Barrat2008}
Alain Barrat, Marc Barthelemy, and Alessandro Vespignani.
\newblock {\em {Dynamical Processes on Complex Networks}}.
\newblock Cambridge University Press, 2008.

\bibitem{Isella2011}
Lorenzo Isella, Juliette Stehl{\'e}, Alain Barrat, Ciro Cattuto, Jean-Fran{\c
  c}ois Pinton, and Wouter Van~den Broeck.
\newblock {What's in a crowd? Analysis of face-to-face behavioral networks}.
\newblock {\em Journal of Theoretical Biology}, 271(1):166--180, 2011.

\bibitem{Miritello2011}
Giovanna Miritello, Esteban Moro, and Rub\'en Lara.
\newblock Dynamical strength of social ties in information spreading.
\newblock {\em Phys. Rev. E}, 83:045102, 2011.

\bibitem{Rocha2010}
Luis E~C Rocha, Fredrik Liljeros, and Petter Holme.
\newblock {Information dynamics shape the sexual networks of Internet-mediated
  prostitution.}
\newblock {\em Proceedings of the National Academy of Sciences of the United
  States of America}, 107(13):5706--5711, 2010.

\bibitem{Bajardi2011}
Paolo Bajardi, Alain Barrat, Fabrizio Natale, Lara Savini, and Vittoria
  Colizza.
\newblock {Dynamical patterns of cattle trade movements}.
\newblock {\em PloS ONE}, 6(5):e19869, 2011.

\bibitem{Butts2009}
Carter~T. Butts.
\newblock Revisiting the foundations of network analysis.
\newblock {\em Science}, 325(5939):414--416, 2009.

\bibitem{Karsai2011}
M.~Karsai, M.~Kivel\"a, R.~K. Pan, K.~Kaski, J.~Kert\'esz, A.-L. Barab\'asi,
  and J.~Saram\"aki.
\newblock Small but slow world: How network topology and burstiness slow down
  spreading.
\newblock {\em Phys. Rev. E}, 83:025102, 2011.

\bibitem{Holme2012}
Petter Holme and Jari Saram{\"a}ki.
\newblock {Temporal networks}.
\newblock {\em Physics Reports}, 519(3):97 -- 125, 2012.

\bibitem{Rocha2013}
Luis E.~C. Rocha and Vincent~D. Blondel.
\newblock Bursts of vertex activation and epidemics in evolving networks.
\newblock {\em PLoS Comput Biol}, 9(3):e1002974, 2013.

\bibitem{Ferreri2014}
Luca Ferreri, Paolo Bajardi, Mario Giacobini, Silvia Perazzo, and Ezio
  Venturino.
\newblock Interplay of network dynamics and heterogeneity of ties on spreading
  dynamics.
\newblock {\em Phys. Rev. E}, 90:012812, 2014.

\bibitem{Iribarren2009}
Jos\'e~Luis Iribarren and Esteban Moro.
\newblock Impact of human activity patterns on the dynamics of information
  diffusion.
\newblock {\em Phys. Rev. Lett.}, 103:038702, 2009.

\bibitem{Vazquez2007}
Alexei Vazquez, Bal\'azs R\'acz, Andr\'as Luk\'acs, and Albert-L\'aszl\'o
  Barab\'asi.
\newblock Impact of non-poissonian activity patterns on spreading processes.
\newblock {\em Phys. Rev. Lett.}, 98:158702, 2007.

\bibitem{Eames2004}
Ken~T.D. Eames and Matt~J. Keeling.
\newblock Monogamous networks and the spread of sexually transmitted diseases.
\newblock {\em Mathematical Biosciences}, 189(2):115 -- 130, 2004.

\bibitem{Perra2012}
Nicola Perra, Bruno Gon{\c c}alves, Romualdo Pastor-Satorras, and Alessandro
  Vespignani.
\newblock {Activity driven modeling of time varying networks}.
\newblock {\em Scientific Reports}, 2:469, 2012.

\bibitem{Volz2009}
Erik Volz and Lauren~Ancel Meyers.
\newblock Epidemic thresholds in dynamic contact networks.
\newblock {\em Journal of The Royal Society Interface}, 6(32):233--241, 2009.

\bibitem{Gross2006}
Thilo Gross, Carlos J.~Dommar D'Lima, and Bernd Blasius.
\newblock Epidemic dynamics on an adaptive network.
\newblock {\em Phys. Rev. Lett.}, 96:208701, 2006.

\bibitem{Taylor2012}
Michael Taylor, Timothy~J. Taylor, and Istvan~Z. Kiss.
\newblock Epidemic threshold and control in a dynamic network.
\newblock {\em Phys. Rev. E}, 85:016103, 2012.

\bibitem{Liu2014}
Suyu Liu, Nicola Perra, M\'arton Karsai, and Alessandro Vespignani.
\newblock Controlling contagion processes in activity driven networks.
\newblock {\em Phys. Rev. Lett.}, 112:118702, 2014.

\bibitem{DeDomenico2013}
Manlio De~Domenico, Albert Sol\'e-Ribalta, Emanuele Cozzo, Mikko Kivel\"a,
  Yamir Moreno, Mason~A. Porter, Sergio G\'omez, and Alex Arenas.
\newblock Mathematical formulation of multilayer networks.
\newblock {\em Phys. Rev. X}, 3:041022, 2013.

\bibitem{Kivela2014}
M.~{Kivel{\"a}}, A.~{Arenas}, M.~{Barthelemy}, J.~P. {Gleeson}, Y.~{Moreno},
  and M.~A. {Porter}.
\newblock {Multilayer Networks}.
\newblock {\em Journal of Complex Networks}, 2014.

\bibitem{VanMieghem2011}
Piet Van~Mieghem.
\newblock The n-intertwined sis epidemic network model.
\newblock {\em Computing}, 93(2-4):147--169, 2011.

\bibitem{Gleeson2012}
James~P. Gleeson, Sergey Melnik, Jonathan~A. Ward, Mason~A. Porter, and
  Peter~J. Mucha.
\newblock Accuracy of mean-field theory for dynamics on real-world networks.
\newblock {\em Phys. Rev. E}, 85:026106, 2012.

\bibitem{VanMieghem2013}
P.~Van~Mieghem and R.~van~de Bovenkamp.
\newblock Non-markovian infection spread dramatically alters the
  susceptible-infected-susceptible epidemic threshold in networks.
\newblock {\em Phys. Rev. Lett.}, 110:108701, 2013.

\bibitem{Li2013}
Cong Li, Huijuan Wang, and Piet Van~Mieghem.
\newblock Epidemic threshold in directed networks.
\newblock {\em Phys. Rev. E}, 88:062802, 2013.

\bibitem{Cator2012}
E.~Cator and P.~Van~Mieghem.
\newblock Second-order mean-field susceptible-infected-susceptible epidemic
  threshold.
\newblock {\em Phys. Rev. E}, 85:056111, 2012.

\bibitem{Mucha2010}
Peter~J Mucha, Thomas Richardson, Kevin Macon, Mason~A Porter, and Jukka-Pekka
  Onnela.
\newblock Community structure in time-dependent, multiscale, and multiplex
  networks.
\newblock {\em Science}, 329:277, 2010.

\bibitem{Wehmuth2014}
Wehmuth Klaus, Ziviani Artur, and Fleury Eric.
\newblock {A unifying model for representing time-varying graphs}.
\newblock {\em arXiv.org}, 2014.

\bibitem{Cozzo2013b}
Emanuele Cozzo, Raquel~A. Ba\~nos, Sandro Meloni, and Yamir Moreno.
\newblock Contact-based social contagion in multiplex networks.
\newblock {\em Phys. Rev. E}, 88:050801, 2013.

\bibitem{Wang2013}
Huijuan Wang, Qian Li, Gregorio D'Agostino, Shlomo Havlin, H.~Eugene Stanley,
  and Piet Van~Mieghem.
\newblock Effect of the interconnected network structure on the epidemic
  threshold.
\newblock {\em Phys. Rev. E}, 88:022801, 2013.

\bibitem{Granell2013}
Clara Granell, Sergio G\'omez, and Alex Arenas.
\newblock Dynamical interplay between awareness and epidemic spreading in
  multiplex networks.
\newblock {\em Phys. Rev. Lett.}, 111:128701, 2013.

\bibitem{Elaydi}
Saber Elaydi.
\newblock {\em An introduction to difference equations}.
\newblock Springer, New York, NY, USA, third edition, 2005.

\bibitem{Lentz2013}
Hartmut H~K Lentz, Thomas Selhorst, and Igor~M Sokolov.
\newblock Unfolding accessibility provides a macroscopic approach to temporal
  networks.
\newblock {\em Phys. Rev. Lett.}, 110:118701, 2013.

\bibitem{Masuda2013}
Naoki Masuda, Konstantin Klemm, and Victor~M Eguiluz.
\newblock {Temporal Networks: Slowing Down Diffusion by Long Lasting
  Interactions}.
\newblock {\em Phys. Rev. Lett.}, 111:188701, 2013.

\bibitem{Erdos1960}
Paul Erd{\H o}s and Alfr{\'e}d R{\'e}nyi.
\newblock {On the evolution of random graphs}.
\newblock {\em Magyar Tud Akad Mat Kutat{\'o} Int K{\"o}zl}, 1960.

\bibitem{Salathe2010}
Marcel Salath{\'e}, Maria Kazandjieva, Jung~Woo Lee, Philip Levis, Marcus~W
  Feldman, and James~H Jones.
\newblock {A high-resolution human contact network for infectious disease
  transmission.}
\newblock {\em Proceedings of the National Academy of Sciences of the United
  States of America}, 107(51):22020--22025, 2010.

\bibitem{Ferreira2012}
Silvio~C Ferreira, Claudio Castellano, and Romualdo Pastor-Satorras.
\newblock {Epidemic thresholds of the susceptible-infected-susceptible model on
  networks: a comparison of numerical and theoretical results.}
\newblock {\em Phys. Rev. E}, 86(4 Pt 1):041125, 2012.

\bibitem{RajKumar2011}
Raj~Kumar Pan and Jari Saram\"aki.
\newblock Path lengths, correlations, and centrality in temporal networks.
\newblock {\em Phys. Rev. E}, 84:016105, 2011.

\bibitem{powell2011}
P~D Powell.
\newblock {Calculating Determinants of Block Matrices}.
\newblock {\em ArXiv e-prints}, 2011.

\bibitem{Binder2010}
Kurt Binder and Dieter~W Heermann.
\newblock {\em {Monte Carlo Simulation in Statistical Physics}}.
\newblock An Introduction. Springer, 2010.

\end{thebibliography}
\bibliographystyle{unsrt}

\newpage

\begin{table}[!h]
\begin{center}
\begin{tabular}{rccccc} 
 network & \# nodes & period T & aggregating window\\
 \hline
 \hline
\ntag{er} \cite{Erdos1960}                   & 500     & 13  &  - \\
\ntag{activity} \cite{Perra2012}            & 1000   & 20  & - \\
\ntag{bursty} \cite{Rocha2013}           & 500     & 50  & - \\
  \hline
\ntag{ht09} \cite{Isella2011}                & 113     & 30    & 1 hour     \\
\ntag{school} \cite{Salathe2010}        & 787     & 42    & 10 mins   \\
\ntag{sex} \cite{Rocha2010}               & 6866   & 13   &  28 days   \\
\end{tabular}
\end{center}
\caption{\label{tab:networks}
{\bf Temporal networks considered for the validation}. The first three networks are single realizations obtained from synthetic models for time-varying networks, the other three are empirical networks. The \ntag{er} model is a sequence of random graphs with 500 nodes and 750 edges, so that $\left\langle k \right\rangle = 3$. The \ntag{activity} model is a sequence of snapshots built with parameters values: $\Delta t = 1$, $m=2$, $\eta=10$, $\gamma = 2.8$, $\epsilon = 3\cdot 10^{-2}$, in notation of \cite{Perra2012}. The \ntag{bursty} network is built with a power law distributed inter-activation time, with exponent $-2$, and cutoff equal to the period of the network.
For the real networks, the collection time is the total time considered in the dataset.
}
\end{table}

\newpage

\begin{figure}[!h]
\begin{center}
 \includegraphics[width=10cm]{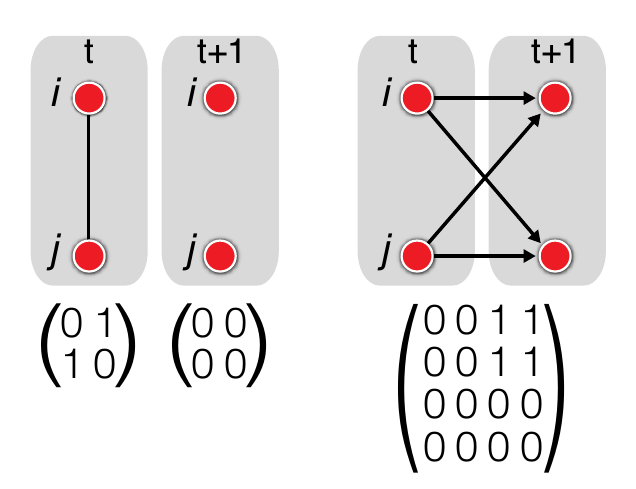}
 \end{center}
\caption{\label{fig:schema}
{\bf Schematic example of the supra-adjacency matrix of the multilayer representation of the temporal network.}
For simplicity, we consider a network of two nodes $i,j$, and two time steps. The left panel represents the network as a sequence of static adjacency matrices. This is translated into a multilayer representation (right panel), where each node points to itself in the future and to the future image of its present neighbors. 
}
\end{figure}

\newpage

\begin{figure}[!h]
\begin{center}
 \includegraphics[width=10cm]{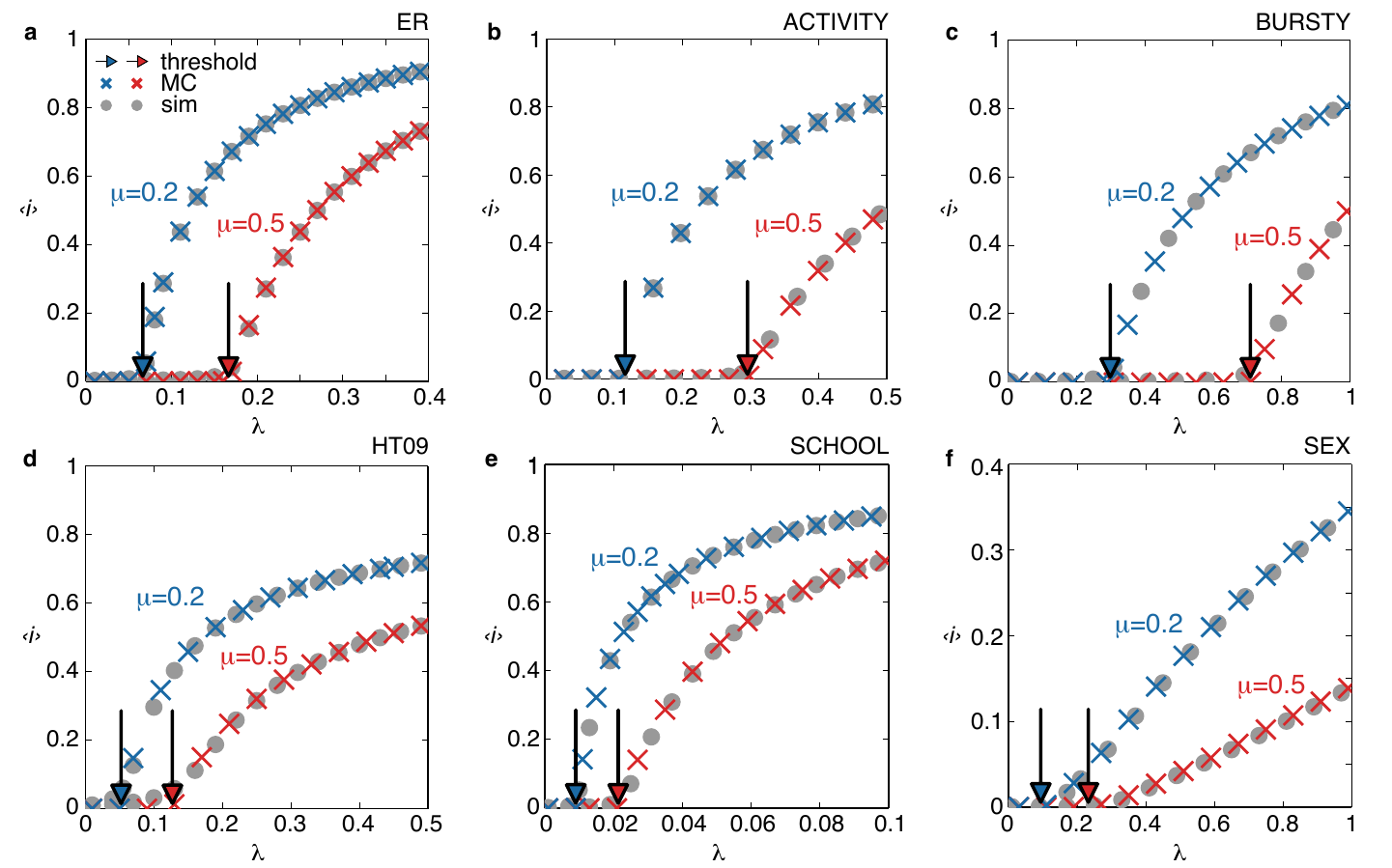}
 \end{center}
\caption{\label{fig:validation}
{\bf Validation of the analytical method and comparison with microscopic numerical simulations.}
Top panel: network models; bottom panel: empirical networks. Cross symbols represent $\langle i_{\text{MC}} \rangle$ as a function of $\lambda$ for two different values of $\mu$ ($\mu=0.2$ in blue and $\mu=0.5$ in red), i.e. the average prevalence obtained from the numerical solution of the Markov chain Eq.~(\ref{eq:temporal}). Circles represent $\langle i_{\text{sim}} \rangle$, i.e. the average prevalence obtained from stochastic microscopic numerical simulations of the SIS process, for the same values of $\mu$. The arrows indicate the analytical predictions of the threshold from Eq.~(\ref{eq:detM}) (one single realization of the network models is considered in each panel).
}
\end{figure}

\newpage

\begin{figure}[!h]
\begin{center}
 \includegraphics[width=10cm]{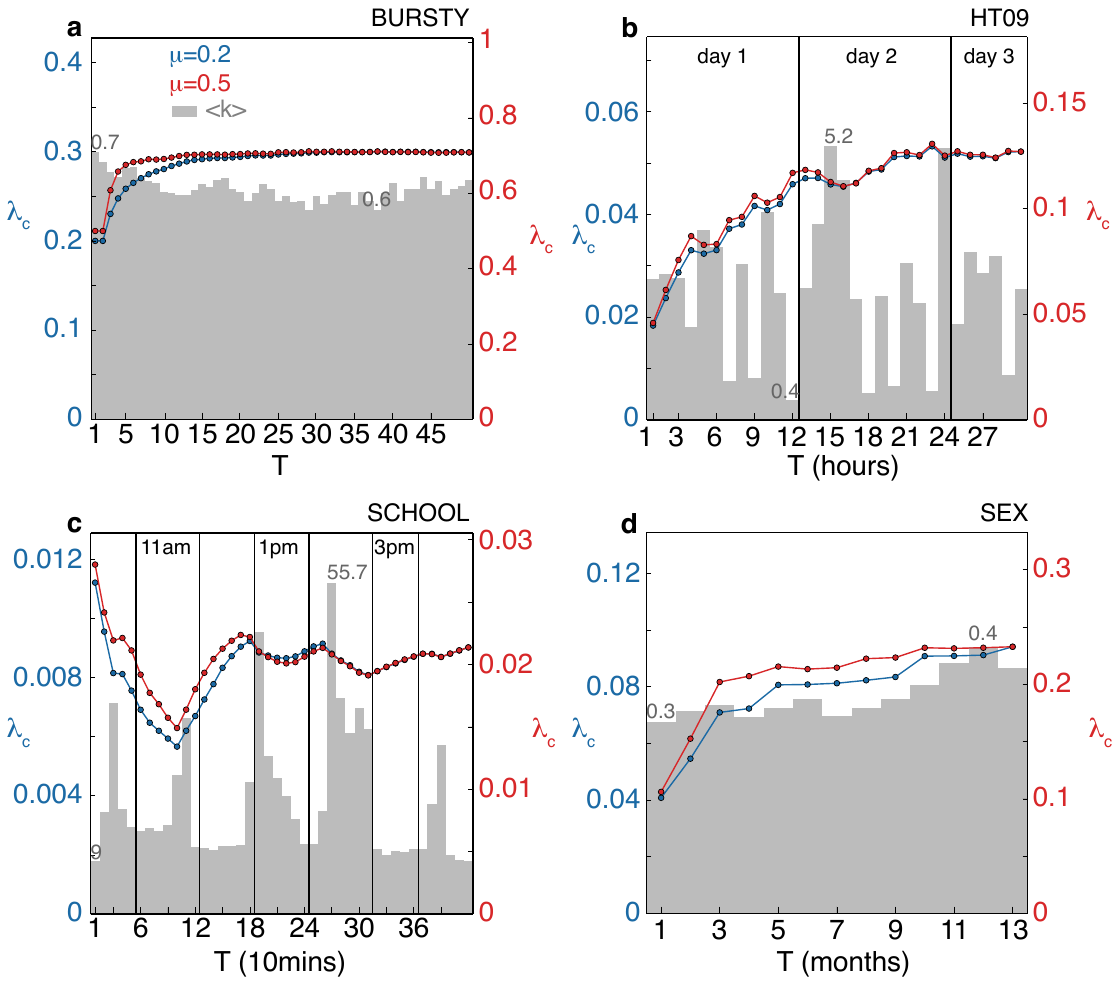}
 \end{center}
\caption{\label{fig:betac_T}
{\bf Epidemic threshold estimated from different period lengths.}
$\lambda_c(T)$ is the epidemic threshold computed by considering only the first $T$ snapshots of the network. For each panel, the blue (red) curve corresponds to $\mu=0.2$ ($\mu=0.5$) and its scale is reported on the left (right) side of the plot. The gray bar chart shows the mean network degree associated to the snapshot at time $T$. Bar charts are in linear scale and min and max values are placed on the corresponding bars. For the empirical networks, the measure of the real time is also reported.
}
\end{figure}

\newpage
\renewcommand\thefigure{S\arabic{figure}}
\setcounter{figure}{0}
\renewcommand\thesection{S\arabic{section}}
\setcounter{section}{0}
\newpage

\begin{center}\Huge
 Supplementary Information
\end{center}
\pagestyle{plain}
\setcounter{page}{1}

\vspace{2cm}

\section{Optimal data collection time} 
\subsection{\ntag{er} and \ntag{activity}}
\begin{figure}[h!]
\begin{center}
 \includegraphics[width=12cm]{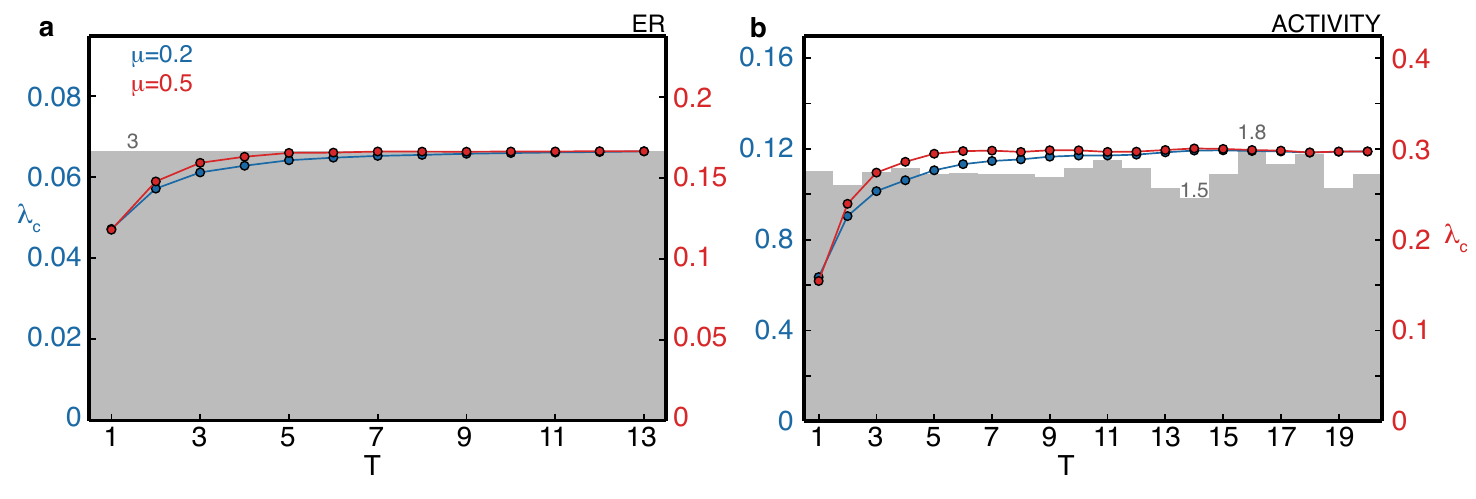}
\end{center}
\caption{
{\bf Epidemic threshold estimated from different period lengths.} $\lambda_c(T)$ is the epidemic threshold computed by considering only the first $T$ snapshots of the network. For each panel, blue (red) curve corresponds to $\mu=0.2$ ($\mu=0.5$). The scale of the former is on the left side of the plot, while the scale of the latter is on the right. The gray bar chart shows the mean network degree associated to the snapshot at time $T$. Bar charts are in linear scale and min and max values are placed on the corresponding bars.
}
\label{fig:betacT_altre}
\end{figure}

Fig.~\ref{fig:betacT_altre} completes the picture given in Fig.~3 of main paper, by showing how period length $T$ affects the epidemic threshold for the models \ntag{er} and \ntag{activity}. In both cases the threshold converges very quickly to a constant value, so that for $T>5$ there are no significant oscillations. This is associated with low variations in the average degree among the different snapshots (variations being exactly equal to zero in the \ntag{er} network by constructions). A small number of snapshots needs to be generated in order to correctly compute the epidemic threshold.
\subsection{More on the correlation between threshold and degree fluctuations}
\begin{figure}[h!]
\begin{center}
 \includegraphics[width=7cm]{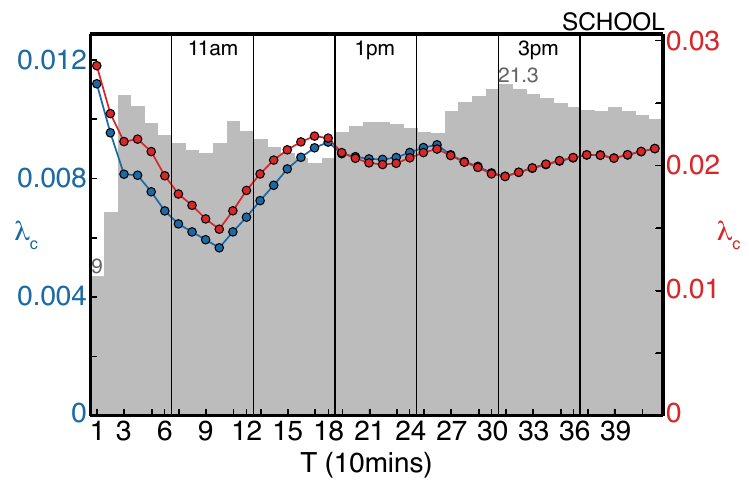}
\end{center}
\caption{
{\bf Epidemic threshold estimated from different period lengths.} $\lambda_c(T)$ is the epidemic threshold computed by considering only the first $T$ snapshots of the network. For each panel, blue (red) curve corresponds to $\mu=0.2$ ($\mu=0.5$). The scale of the former is on the left side of the plot, while the scale of the latter is on the right. The gray bar chart shows the average of the mean network degree of snapshots up to time $T$. Bar charts are in linear scale and min and max values are placed on the corresponding bars. The measure of the real time is also reported. 
}
\label{fig:betacT_meanmean}
\end{figure}

In main paper we show how the oscillations of $\lambda_c$ are associated with the instantaneous fluctuations of the average degree of the network, as the period $T$ varies. As $T$ increases, $\lambda_c$ fluctuations are damped, because they are cumulatively calculated on longer periods. Adding a snapshot to a short period $T$ has indeed a greater relative influence on the threshold than adding it to an already long period. Similarly, damped oscillations are observed when we monitor the cumulative average degree of the network, calculated on all snapshots up to period $T$ (Figure~\ref{fig:betacT_meanmean}).
\section{Higher order correlations}
\begin{figure}[h!]
\begin{center}
 \includegraphics[width=8cm]{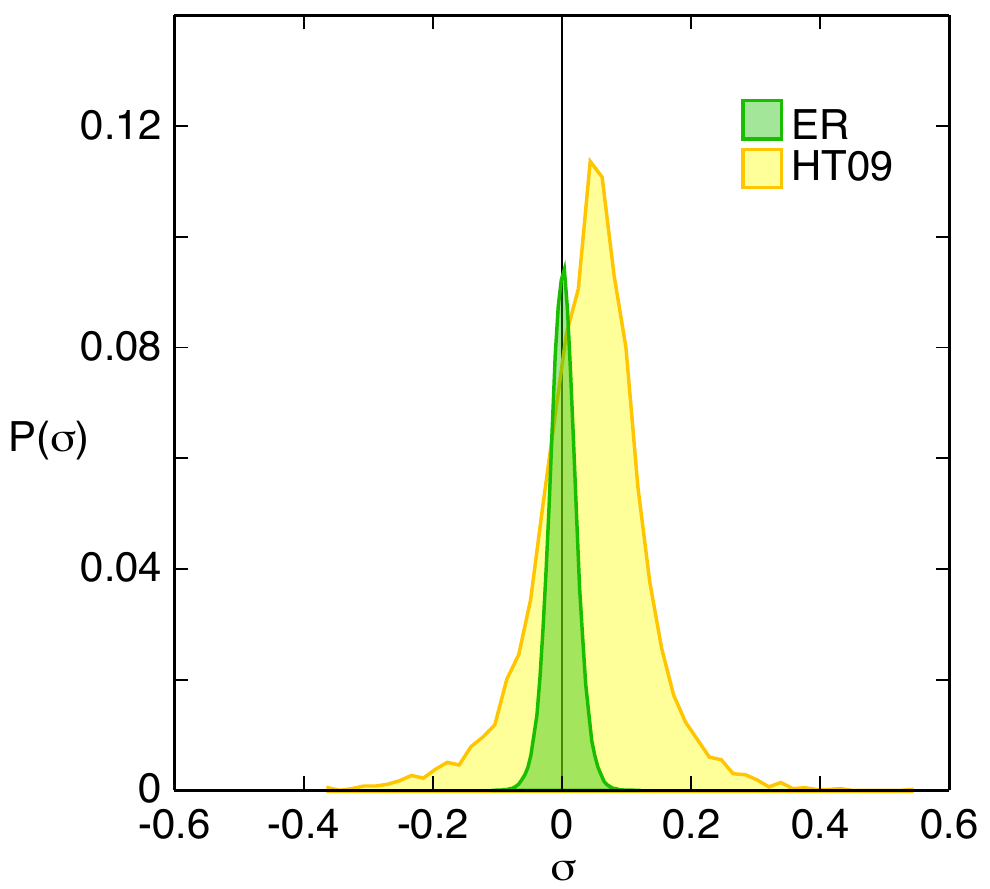}
\end{center}
\caption{
{\bfseries Two-point correlations for \ntag{er} and \ntag{ht09}.} Distribution of $\sigma=\sigma_{ij}$ for the model \ntag{ER} and the real network \ntag{HT09}. Distributions are computed using $\lambda=0.3$, $\mu=0.5$ for both networks. Moments are computed over $2\cdot 10^5$ time steps, after a relaxation time of $3\cdot 10^3$.
}
\label{fig:2point}
\end{figure}

The quenched mean field approach disregards spatial correlations among the probabilities of nodes being infected. Let $X_i(t)=0,1$ be the infectious status of node $i$ at time $t$. The quenched mean field then assumes that $\left\langle X_iX_j\right\rangle = \left\langle X_i\right\rangle\left\langle X_j\right\rangle$. In order to assess the impact of such approximation, at least in the case of two point correlations, we compute in the simulations the following quantity for each pair of nodes:
\begin{equation}
 \sigma_{ij} = \frac{\left\langle X_iX_j\right\rangle - \left\langle X_i\right\rangle\left\langle X_j\right\rangle}{\sqrt{\left\langle X_i^2\right\rangle-\left\langle X_i\right\rangle^2}  \sqrt{\left\langle X_j^2\right\rangle-\left\langle X_j\right\rangle^2}   },
\end{equation}
where the moments are computed on a time interval well after the initial transient: $\left\langle Y\right\rangle = \left[\sum_{t=t_0}^{t_1}Y(t)\right]/\left(t_1-t_0\right)$.
$\sigma_{ij}\approx 1$ indicates that $X_i,X_j$ are highly correlated, while low values of that quantity indicate no correlations.
Figure~\ref{fig:2point} shows the distribution for $\sigma_{ij}$ on all possible pairs of nodes, for \ntag{er} and \ntag{ht09}. For \ntag{er} almost no correlations are visible among nodes, as expected, except for small fluctuations around $\sigma_{ij}=0$. For \ntag{ht09}, on the other hand, we see that $\sigma_{ij}$ is peaked around a small but nonzero value, indicating the presence of two point correlations, albeit weak. This is in accordance with the finding that the threshold computation is more accurate for \ntag{er} than for \ntag{ht09} (see Figure~2a,d of main paper).
\section{Weighted and directed networks}
For sake of simplicity, in the main paper we deal with undirected unweighted temporal networks, i.e. $A^{(t)\dagger}=A^{(t)}$ and $A^{(t)}_{ij}\in \left\{ 0,1 \right\}$. Here we briefly show that our methodology can be extended to the more general case, if needed.

The proposed approach does not require $A^{(t)}$ to be symmetric (moreover $M_{\alpha\beta}$ is not symmetric even when the $A^{(t)}$ matrices are). Therefore a generalisation to the directed case is straightforward, and it is simply based on the replacement of ${\mathcal A}$ with ${\mathcal A}^\dagger$ in the computation of the threshold in the annealed approximation.

The weighted case is tractable too, provided that the probability of transmitting the infection is defined in terms of the weight, according to the given context under study (the weight could for example represent the movements of hosts from one node to another).

It is customary to compute the probability of transmission along a link $ij$ as $1-\left(1-\lambda\right)^{A_{ij}^{(t)}}$ (see, for instance, [1]). When the weights are integer numbers, this means considering the binomially distributed probability of at least one infection given $A_{ij}^{(t)}$ trials. By plugging this contribution into $\mathbf{M}_{ij}^{tt'}$, equation~(4) of the main paper thus becomes
\begin{equation}
  \mathbf{M}_{ij}^{tt'} = \delta^{t,t'+1}\left[  \left(1-\mu\right)\delta_{ij} + 1-\left(1-\lambda\right)^{A_{ij}^{(t)}}  \right].
\end{equation}
The computation of the threshold can then be carried out in the same way as described in main paper.

\section{Computational performance}
In this section we discuss the performance and scalability of the numerical algorithm we use to compute the spectral radius of matrix $P$. The algorithm is implemented in \texttt{Python 2.7} and uses \texttt{numpy} and \texttt{scipy} libraries for sparse matrix representation and multiplication, \texttt{numpy.dot} and \texttt{scipy.sparse.csr}. The spectral radius of $P$ is computed through a modified version of the well known power iteration method [2]. We tested the scaling of the execution time by varying the number of nodes $N$ and the period $T$ of an uncorrelated sequence of Erd\H{o}s-R\'eny with $\langle k\rangle=2$. The execution time grows linearly with the period and as the $5/2$-th power of the number of nodes, $t\sim TN^{5/2}$. As a point of reference, the average execution time for $N=10^3$ and $T=10^2$ is $15$ seconds on an Intel Core i5 3.2G Hz with RAM DDR3 $8$ GB and $1.6$ GHz.

\vspace{1cm}

\begin{flushleft}
 {\large\bf SI References}\\
 
 \vspace{0.4cm}
 
 [1] Bajardi, P., Barrat, A., Savini, L. \& Colizza, V. Optimizing surveillance for livestock disease spreading through animal movements. {\itshape J. R. Soc. Interface}. doi:10.1098/rsif.2012.0289 (2012).
 
 [2] M\"untz, H, Solution directe de l'\'equation s\'eculaire et de quelques probl\`emes analogues transcendants, {\itshape C. R. Acad. Sci. Paris} 156,43-46 (1913).
 
\end{flushleft}

\end{document}